\documentclass{jnmp}
\usepackage{amsmath}

\begin{document}
\setcounter{page}{288}
\newtheorem{lemma}{Lemma}[section]
\newtheorem{proposition}{Proposition}[section]
\newtheorem{corollary}{Corollary}[section]
\newtheorem{theorem}{Theorem}[section]

\renewcommand{\evenhead}{J van de Leur}
\renewcommand{\oddhead}{Matrix Integrals and the Geometry of Spinors}

\thispagestyle{empty}


\FirstPageHead{8}{2}{2001}
{\pageref{Leur-firstpage}--\pageref{Leur-lastpage}}{Article}

\copyrightnote{2001}{J van de Leur}

\Name{Matrix Integrals and the Geometry of Spinors}\label{Leur-firstpage}

\Author{Johan VAN DE LEUR}

\Address{Mathematical Institute, Utrecht University\\
P.O. Box 80010, 3508 TA Utrecht, The Netherlands\\
E-mail: vdleur@math.uu.nl}

\Date{Received October 9, 2000; Accepted January 10, 2001}

\begin{abstract}
\noindent
We obtain the collection of symmetric and symplectic matrix integrals and the
collection of
Pfaffian tau-functions, recently described by Peng and Adler and van Moerbeke,
as specific elements in the Spin-group orbit of the vacuum vector of a
fermionic Fock space. This fermionic Fock space is the same space as one
constructs to obtain the KP and 1-Toda lattice hierarchy.
\end{abstract}

\rightline{\bfseries\itshape In memory of F~D~Veldkamp}

\renewcommand{\theequation}{\thesection.\arabic{equation}}
\setcounter{equation}{0}

\section{Introduction}
Recently H~Peng \cite{leur:Pe} showed that the symmetric matrix model and the
statistics of the spectrum of symmetric matrix ensembles is governed by a
strange reduction of the 2-Toda lattice hierarchy~\cite{leur:UT}. Adler  and van
Moerbeke
\cite{leur:Pfaf,leur:Pfaf2} (see also~\cite{leur:ASvM})
show that this reduction leads to vectors of Pfaffian
tau-functions. However these  Pfaffian tau-functions do not satisfy the Toda
lattice hierarchy, but rather another system of PDE's, which can be identified
as the BKP hierarchy in the form described by V~Kac and the author in~\cite{leur:KvdL}.
To be more explicit, let ${\cal S}_{m}(E)$ be the set of $m\times m$
symmetric matrices with spectrum in $E\subset\mathbb{R}$ (a union of
intervals), $dZ$ the Haar measure on symmetric matrices and
$V(t,z)=V(z)+\sum\limits_{i=1}^\infty
t_iz^i$, where $V(Z)$ is a potential.
Applying the spectral theorem to the symmetric matrix $Z=
O^{T} \mbox{diag}\left(z^0,...,z_{m-1}\right)O$, with $O\in SO(m)$, we find
upon integrating over the special orthogonal group the following formula for
the symmetric matrix integral  (see also~\cite{leur:Pfaf}).
\begin{equation}\label{leur:0.1}
\hat\tau_{m}^E(t)=\int_{{\cal S}_{m}(E)}e^{Tr V(t,Z) } dZ
=c_{m}\int_{\mathbb{R}^{m}}|\Delta_{m}(z)|\prod_{i=0}^{m-1}(e^{V(t,z_i)}
I_E(z_i)dz_i),
\end{equation}
where $\Delta_{m}(z)=\prod\limits_{0\le i<\le j\le m-1} (z_j-z_i)$ is the
Vandermonde
 determinant. Adler and van Moerbeke  \cite{leur:Pfaf,leur:Pfaf2}
 show (see also Section~2) that these integrals for $m=2n$ can be expressed in certain
Pfaffians:
\begin{equation}
\label{leur:0.2}
\hat\tau^E_{2n}
=(2n)!c_{2n}\text{Pf}\left( (\mu_{k,\ell}(t))_{0\le k,\ell\le 2n-1}\right ).
\end{equation}
Here $\mbox{Pf}(A)$ stands for the Pfaffian of a skew-symmetric
matrix $A=(A_{ij})_{0\le i,j \le 2n-1}$ defined by
\[
\mbox{\rm Pf}(A)=\frac{1}{2^nn!}\sum_{\sigma\in S_{2n}}\text{sg}(\sigma)
\prod_{j=0}^{n-1 }
A_{\sigma(2j),\sigma(2j+1)}.
\]
(if $n=0$ we assume Pf$(A)=1$)
and $\mu_{ij}(t)=\langle y^i,z^j\rangle_t$
are the moments of the time-dependent skew-symmetric inner product
\begin{equation}
\label{leur:0.3}
\langle f,g\rangle_t=\iint_{E^2}f(y)g(z)e^{V(t,y)+V(t,z)}\text{sg}(z-y)dydz.
\end{equation}
The formula for $\hat\tau^E_{m}$ will be obtained Section~2.
If $m$ is odd, it is also a Pfaffian, but the expression is more complicated.

In this paper we will show, using the Clifford algebra techniques of
\cite{leur:KvdL}, that replacing $\hat\tau_{m}^E$ by
$\tau_{m}^E=\frac{\hat\tau_{m}^E}{m!c_{m}}$,  the generating series,
\[
\tau^E(t,q)=\sum_{m=0}^\infty \tau_{m}^E(t)q^{m},
\]
of these Pfaffians is a specific element in the Spin group orbit of the vacuum
vector in a fermionic Fock space. The fermionic Fock space which is constructed
is more or less the same as the one one uses to obtain the KP hierarchy and
1-dim Toda
lattice hierarchy. However, these tau-functions do not satisfy the KP or Toda
lattice hierarchy, but we will show in section 4 that they satisfy the
(charged) BKP hierarchy of~\cite{leur:KvdL} (see also Section~3):
\begin{equation}\label{leur:0.12}
\arraycolsep=0em
\begin{array}{l}
\mbox{Res}\; \Big (z^{n-1} e^{\xi(t',z)}e^{-\eta(t',z)}\tau_{n-1}(t')
z^{-m-1} e^{-\xi(t'',z)}e^{\eta(t'',z)}\tau_{m+1}(t'')
\vspace{2mm}\\
\qquad{}+(z)^{-n-1} e^{-\xi(t',z)}e^{\eta(t',z)}\tau_{n+1}(t')(z)^{m-1}
e^{\xi(t'',z)}e^{-\eta(t'',z)}\tau_{m-1}(t'')\Big )
\vspace{2mm}\\
\displaystyle \qquad {} =\frac{1}{2}\left (1-(-1)^{n+m}\right )\tau_n(t')\tau_m(t''),
\end{array}
\end{equation}
for all $n,m\in\mathbb Z$, where $\mbox{Res}\;\sum f_iz^i=f_{-1}$ and
\begin{equation}
\label{leur:0.4}
\xi (t,z) = \sum^{\infty}_{k=1} t_k z^k \qquad \mbox{and}\qquad \eta(t,z)
= \sum^{\infty}_{k=1} \frac{1}{k} \frac{\partial}{\partial t_k} z^{-k} .
\end{equation}

Note that for all $n,m\in 2\mathbb{Z}$ these are exactly the equations for
the Pfaffian tau-functions obtained by Adler and van Moerbeke in \cite{leur:Pfaf}.
The hierarchy of equations for all  $n,m\in 2\mathbb{Z}$ is called the DKP
hierarchy in \cite{leur:KvdL}.

In fact using the KP boson-fermion correspondence one shows
that
\[\arraycolsep=0em
\begin{array}{l}
\displaystyle \tau^E(t,q)=\exp \left
(\frac{1}{2}\iint_{E^2}X(t,y,z)e^{V(y)+V(z)}\text{sg}(z-y)dydz\right )
\vspace{3mm}\\
\displaystyle \phantom{\tau^E(t,q)=}{}\times
\left( 1+\int_EX(t,w)e^{V(w)}dw\right )\cdot 1,
\end{array}
\]
where $X(t,z)$ and $X(t,y,z)$ are  the vertex operators
\begin{equation}
\label{leur:X}
\arraycolsep=0em
\begin{array}{l}
\displaystyle X(t,z)=qz^{q\frac{\partial}{\partial q}}
\exp(\xi (t,z))
\exp(-\eta(t,z)),
\vspace{2mm}\\
\displaystyle X(t,y,z)=X(t,z)X(t,y)
\vspace{2mm}\\
\phantom{X(t,y,z)}{}=q^2(z-y)(yz)^{q\frac{\partial}{\partial q}}
\exp(\xi (t,z)+\xi (t,y))
\exp(-\eta(t,z)-\eta(t,y)).
\end{array}
\end{equation}

The symplectic matrix integrals, which can be treated in a similar way,
are described in Section~8.

\setcounter{equation}{0}

\section{Symmetric matrix integrals}
Let from now on $m=2n$ if
$m$ is even and $m=2n+1$ if $m$ is odd and recall formula (\ref{leur:0.1}):
\[
\hat\tau_{m}^E(t)=
c_{m}\int_{\mathbb{R}^{m}}|\Delta_{m}(z)|\prod_{i=0}^{m-1}\left(e^{V(t,z_i)}
I_E(z_i)dz_i\right).
\]
Denote by  sg  the sign-function and by $S_{m}$  the permutation
group of $m$ letters. Using the identity
\[
\sum_{\sigma\in S_{m}}\text{sg}(\sigma)\prod_{j=0}^{n
-1}\text{sg}(\sigma(2j+1)-\sigma(2j))=2^nn!,
\]
we find
\begin{equation}\label{leur:P1.1}
\begin{aligned}
 \hat\tau_{m}^E(t)
={}& m!c_{m}\int_{-\infty<z_0<z_1<\ldots
<z_{m-1}<\infty}\Delta_{m}(z)\prod_{i=0}^{m-1}\left(e^{V(t,z_i)}
I_E(z_i)dz_i\right)\\
={}& \frac{m!c_{m}}{2^nn!}
\sum_{\sigma\in S_{m}}\text{sg}(\sigma)\prod_{j=0}^{n
-1}\text{sg}\left(\sigma^{-1}(2j+1)-\sigma^{-1}(2j)\right)\\
 & \qquad \times \int_{-\infty<z_0<z_1<\ldots
<z_{m-1}<\infty}\Delta_{m}(z)\prod_{i=0}^{m-1}\left(e^{V(t,z_i)}
I_E(z_i)dz_i\right)\\
 ={}& \frac{m!c_{m}}{2^nn!}
\sum_{\sigma\in S_{m}}\text{sg}(\sigma)\int_{-\infty<z_0<z_1<\ldots
<z_{m-1}<\infty}\Delta_{m}(z)\\
&\qquad \times\prod_{j=0
}^{n-1} \text{sg}\left(z_{\sigma^{-1}(2j+1)}-z_{\sigma^{-1}(2j)}\right)
\prod_{i=0}^{m-1}\left(e^{V(t,z_i)} I_E(z_i)dz_i\right)\\
 ={} & \frac{m!c_{m}}{2^nn!}
\sum_{\sigma\in S_{m}}\int_{-\infty<z_{\sigma(0)}<z_{\sigma(1)}
<\ldots <z_{\sigma(m-1)}<\infty}\Delta_{m}(z)\\
&\qquad \times \prod_{j=0
}^{n-1} \text{sg}(z_{2j+1}-z_{2j})\prod_{i=0}^{m-1}\left(e^{V(t,z_i)} I_E(z_i)dz_i\right)\\
 ={}& \frac{(m)!c_{m}}{2^nn!}
\int_{\mathbb{R}^{m}}\Delta_{m}(z)\prod_{j=0}^{n-1}\text{\rm
sg}(z_{2j+1}-z_{2j})\prod_{i=0}^{m-1}\left(e^{V(t,z_i)}
I_E(z_i)dz_i\right).
\end{aligned}
\end{equation}
If $m=2n$ is even we continue as follows:
\begin{equation}
\label{leur:P1.1a}
\begin{aligned}
 \hat\tau^E_{2n}={}& \frac{(2n)!c_{2n}}{2^nn!}
\sum_{\sigma\in S_{2n}}\text{sg}(\sigma)
\prod_{j=0}^{n-1 }\int_{\mathbb{R}^{2}}
\Bigl(z_{2j}^{\sigma(2j)}z_{2j+1}^{\sigma(2j+1)}
e^{V(t,z_{2j})+V(t,z_{2j+1})}\\
&\qquad \times I_E(z_{2j})I_E(z_{2j+1})
\text{sg}(z_{2j+1}-z_{2j})dz_{2j}dz_{2j+1}\Bigr)\\
={}&\frac{(2n)!c_{2n}}{2^nn!}
\sum_{\sigma\in S_{2n}}\text{sg}(\sigma)\prod_{j=0}^{n-1 }
\mu_{\sigma(2j),\sigma(2j+1)}(t)
=(2n)!c_{2n}\text{Pf}\left( (\mu_{k,\ell}(t))_{0\le k,\ell\le 2n-1}\right ).
\end{aligned}\hspace{-5mm}
\end{equation}
This is formula (\ref{leur:0.2}) of the introduction, with the moments
$\mu_{ij}$ defined by (\ref{leur:0.3}).

If $m=2n+1$ is odd, we use the following lemma, which can be found in Adler,
Horozov and van Moerbeke~\cite{leur:AHvM}, to continue the calculation.

\begin{lemma}\label{leur:L1.1}
Let $A=(a_{ij})_{0\le i,j\le 2\ell+1}$ be a skew symmetric matrix, then
\[\text{\rm Pf}(A)=\sum_{k=0}^{2\ell} (-)^k
A_{k,2\ell+1}\text(Pf)((A_{ij})_{i,j\ne k,2\ell+1}).
\]
\end{lemma}

Now,
\begin{equation}
\label{leur:P1.1b}
\begin{aligned}\hat\tau^E_{2n+1}(t)
={}&\frac{(2n+1)!c_{2n+1}}{2^nn!}
\sum_{\sigma\in S_{2n+1}}\text{sg}(\sigma)
\int_{\mathbb{R}^{2n+1}}\prod_{j=0}^{n-1}\text{\rm sg}(z_{2j+1}-z_{2j})\\
& \qquad \times
\prod_{i=0}^{2n}z_i^{\sigma(i)}\left(e^{V(t,z_i)}
I_E(z_i)dz_i\right)\\
={}&\frac{(2n+1)!c_{2n+1}}{2^nn!}
\sum_{k=0}^{2n}
\sum_{\sigma\in S_{2n+1},\sigma(2n)=k}\text{sg}(\sigma)
\int_{\mathbb{R}^{2n+1}}z_{2n}^ke^{V(t,z_{2n})}
I_E(z_{2n})dz_{2n}\\
\ &\qquad\times
\prod_{j=0}^{n-1}\text{\rm sg}(z_{2j+1}-z_{2j})
\prod_{i=0}^{2n-1}z_i^{\sigma(i)}\left(e^{V(t,z_i)}
I_E(z_i)dz_i\right)\\
={}&\frac{(2n+1)!c_{2n+1}}{2^nn!}
\sum_{k=0}^{2n}(-)^k
\sum_{\rho\in S_{2n}^{(k)}}\text{sg}(\rho)
\int_{\mathbb{R}^{2n+1}}z_{2n}^ke^{V(t,z_{2n})}
I_E(z_{2n})dz_{2n}\\
\ &\qquad\times
\prod_{j=0}^{n-1}\text{\rm sg}(z_{2j+1}-z_{2j})
\prod_{i=0}^{2n-1}z_i^{\rho(i+\epsilon_k(i))}\left(e^{V(t,z_i)}
I_E(z_i)dz_i\right),
\end{aligned}\hspace{-5mm}
\end{equation}
where  $\rho\in S_{2n}^{(k)}$, the set of permutations of the numbers
$1,2,\ldots,k-1,k+1,\ldots,2n$, is such that $\sigma(j)=\rho(j+\epsilon_k(j))$,
with
$\epsilon_k(j)=0$ if $j< k$ and $=1$ if $j\ge k$. To give an idea of what we
are doing, we give an example. If $n=2$, $k=1$ and
$\sigma=\left(
\begin{array}{ccccc}
0&1&2&3&4\\
4&3&0&2&1
\end{array}\right )
$,
then
$\rho=\left (
\begin{array}{cccc}
0&2&3&4\\
4&3&0&2
\end{array}\right )
$ and sg$(\sigma)=-$sg$(\rho)$.
Hence,

\newpage

\noindent
\begin{equation}
\label{leur:P1.1c}
\begin{aligned}
\hat\tau^E_{2n+1}(t)
={}&\frac{(2n+1)!c_{2n+1}}{2^nn!}
\sum_{k=0}^{2n}(-)^k
\int_{\mathbb{R}}z_{2n}^k\left(e^{V(t,z_{2n})}
I_E(z_{2n}\right)dz_{2n})
\sum_{\rho\in S_{2n}^{(k)}}\text{sg}(\rho)\\
&\qquad\times
\prod_{j=0}^{n-1 }\int_{\mathbb{R}^{2}}
\Bigl(z_{2j}^{\rho(2j+\epsilon_k(2j))}z_{2j+1}^{\rho(2j+1+\epsilon_k(2j+1))}
e^{V(t,z_{2j})+V(t,z_{2j+1})}\\
\ &\qquad\times
I_E(z_{2j})I_E(z_{2j+1})\text{sg}(z_{2j+1}-z_{2j})dz_{2j}dz_{2j+1}\Bigr)\\
={}&(2n+1)!c_{2n+1}
\sum_{k=0}^{2n}(-)^k
\int_{\mathbb{R}}z_{2n}^k\left(e^{V(t,z_{2n})}
I_E(z_{2n})dz_{2n}\right)\\
& \qquad \times \text{Pf}((\mu_{ij}(t))_{0\le i,j\le 2n, i,j\ne k})\\
={}&(2n+1)!c_{2n+1}\sum_{k=0}^{2n}(-)^k
\mu_k(t)\text{Pf}((\mu_{ij}(t))_{0\le i,j\le 2n, i,j\ne k})\\
={}&(2n+1)!c_{2n+1}\text{\rm Pf}(M_{2n+1}(t)),
\end{aligned}
\end{equation}
where
\begin{equation}
\label{leur:M}
M_{2n+1}(t)=\left(
\begin{array}{ccc|c}
 & & &\mu_0(t)\\
 &(\mu_{ij}(t))_{0\le i,j\le 2n}& & \vdots\\
 & & & \vdots\\
 & & &\mu_{2n}(t)\\
\hline
-\mu_0(t)&\cdots\ \cdots&-\mu_{2n}(t)&0
\end{array}\right)
\end{equation}
and $\mu_i(t)=(1,z^i)_t$ are the moments of the time-dependent symmetric
inner product
\begin{equation}
\label{leur:inner}
(f,g)_t=\int_E f(z)g(z)e^{V(t,z)}dz.
\end{equation}
The symplectic matrix integrals, which can be treated in a similar way,
are described in Section 8.

\setcounter{equation}{0}

\section{The geometry of spinors and the BKP hierarchy}
In this section we recall the results of \cite{leur:KvdL}.

Consider the vector space $V=V^+\oplus V^0\oplus V^-$, where
$V^\pm=\bigoplus\limits_{i\in\mathbb{Z}+\frac{1}{2}}\mathbb{C}\psi_i^\pm $
and $V^0=\mathbb{C}\psi_0$
with symmetric bilinear form
\begin{equation}
\label{leur:0.-1}
(\psi_0,\psi_0)=1,\qquad
(\psi_0,\psi^\pm_i)=0,\qquad
(\psi^\pm_i,\psi_j^\pm)=0,\qquad(\psi^\pm_i,\psi_j^\mp)=\delta_{i,-j}.
\end{equation}
Let $C\ell\; V$ be the associated Clifford algebra, that is the quotient
of the tensor algebra over~$V$ by the ideal generated by relations
\begin{equation}
  \label{leur:1.1}
  uv + vu = (u,v)1,\qquad\mbox{where} \ \ u,v \in V.
\end{equation}
These relations induce a natural $\mathbb{Z}/2\mathbb{Z}$ decompostion
$C\ell\; V=C\ell\; V_{\overline{0}}\oplus C\ell\; V_{\overline{1}}$.
Denote by  $(C\ell\; V)^\times$ the multiplicative group of invertible
elements of the algebra $C\ell\; V$, by $\mbox{Pin}\;V$ the subgroup
of $(C\ell\; V)^\times$ generated by all the elements $a$ such that
$aVa^{-1}=V$ and let Spin $V=$ Pin $V\cap C\ell_{\overline{0}}V$.
A simple but for this paper important observation is that for $\lambda=+,-$,
both
$1+\sum\limits_{-N\le i\le M} c_i \psi_0\psi^\lambda_i\in\text{Spin}\;V$
and
\begin{equation}
\label{leur:spin}
\exp\left (\sum_{-N\le i<j\le M}c_{ij}\psi^\lambda_i\psi^\lambda_j\right
)\in\text{\rm Spin}\; V.
\end{equation}

We have a homomorphism $T:$
$\mbox{Pin}\; V\rightarrow O(V), g\mapsto T_g$ defined by $(v\in V)$:
\[
T_g (v)= g v g^{-1}  \in V.
\]
Denote by   $U=\sum\limits_{j>0}(\mathbb{C}\psi_j^++\mathbb{C}\psi_j^-
)+\mathbb{C}(1+\sqrt 2\psi_0)$  the
subspace of $\mathbb{C}1+V$ and let
\[
F:=F(V,U) = C\ell\; V/ (C\ell\; V)U.
\]
The space $F(V,U)$
caries a structure of a $C\ell\; V$-module induced by left
multiplication. This module restricted to $\mbox{Pin}\; V$ is called the {\it
  spin module} of the group $\mbox{Pin}\; V$. This module remains irreducible when
restricted to $\mbox{Spin}\;V$. Denote  the image of $1$ in $F(V,U)$ by
$|0\rangle $, then
\[
\psi_j^\pm|0\rangle=0\quad\mbox{if}\ \ j>0\qquad\text{and}
\qquad \psi_0|0\rangle=-\frac{1}{\sqrt 2}|0\rangle.
\]
By introducing the notion of charge as follows:
\[
\mbox{charge}\;|0\rangle=\mbox{charge}\;\psi_0=0,\qquad
\mbox{charge}\;\psi_j^\pm=\pm 1,
\]
the space $F$ decomposes into charge sectors
\[
F=\bigoplus_{k\in\mathbb{Z}}F_k.
\]
Given $f\in F$, let
\[\text{\rm Ann}\;  f = \{ v\in V \mid v f = 0\},
\]
then the vacuum vector $|0\rangle  \in F$ is characterized (up
to a constant factor) among the vectors of $F$ by the property that
\[
\text{\rm Ann}\;  |0\rangle
= U_0:=\sum_{j>0}(\mathbb{C}\psi_j^++\mathbb{C}\psi_j^- ).
\]
Moreover, let $g\in$ Pin $V$, then
\[\text{\rm Ann}\;g|0\rangle=T_g(\text{Ann} \; |0\rangle)= T_g(U_0).
\]
All maximal isotropic subspaces of $V$ characterize the
$\mbox{Spin}\; V$-group orbit.
Let
 $O=\mbox{Spin}\; V\cdot|0\rangle$
be
the $\mbox{Spin}\; V$-orbit of
$| 0\rangle$,
one of the main observations of the paper~\cite{leur:KvdL} is the following:

\begin{proposition}
\label{leur:t1.8}
 If $\tau \in F$
and $\tau\ne 0$, then $\tau\in
     O$ if and only if $\tau$ satisfies the equation
     \begin{equation}
       \label{leur:1.10}
\psi_0\tau\otimes \psi_0\tau +\sum_{j\in\mathbb{Z}+\frac{1}{2}}
(\psi_j^+\tau\otimes
\psi_{-j}^-\tau
+\psi_j^-\tau\otimes
\psi_{-j}^+\tau)
=\frac{1}{2}\tau\otimes\tau.
\end{equation}
     \end{proposition}

This equation is called the fermionic BKP hierarchy. We will now rewrite these
equations to a hierarchy of differential equations. For this we use
the classical boson-fermion correspondence~\cite{leur:KacRaina}.
Consider the following generating series,  called {\it charged fermionic
fields}:
\begin{equation}
  \label{leur:2.1}
  \psi^{\pm}{(z)} = \sum_{i\in \frac{1}{2} + \mathbb{Z}} \psi^{\pm}_i
  z^{-i -\frac{1}{2}} .
\end{equation}
Then we have:
\begin{equation}
  \label{leur:2.2}
  \psi^{\lambda} (y) \psi^{\mu} (z) + \psi^{\mu} (z) \psi^{\lambda}
  (y) = \delta_{\lambda,-\mu}\delta (y-z),\qquad\lambda,\mu=\pm,
\end{equation}
where
\[
\delta (y-z) = y^{-1} \sum_{n\in \mathbb{Z}} \left(\frac{y}{z} \right)^n.
\]

We split up field $\phi{(z)}=\sum\limits_i \phi_i z^i$ in its positive and negative
part:
\[
\phi{(z)} = \phi(z)_+ + \phi(z)_-,
\]
where
\[
\phi_+ = \sum_{i\ge 0} \phi_i z^i \quad \mbox{and} \quad
\phi(z)_- = \phi(z) - \phi(z)_+ ,
\]
Define the bosonic fields ($\nu\in\mathbb{C}$)
\begin{equation}
  \label{leur:2.3}
\begin{aligned}
{}&  \alpha (z)= \sum_{k\in \mathbb{Z}} \alpha_k z^{-k-1} = : \psi^{+}
    (z) \psi^- (z):,\\
{}& L^\nu(z)=\sum_{k\in\mathbb{Z}} L_k^\nu
z^{-k-2}=\frac{1}{2}:\alpha(z)\alpha(z):+\left(\frac{1}{2}-\nu\right)
\partial_z(\alpha(z)),\\
{}& Y^\pm (z)=\sum_{k\in\mathbb{Z}} Y_k^\pm z^{-k-2}
=\partial_z(\psi^\pm(z))\psi^\pm (z),
\end{aligned}
\end{equation}
where the normally ordered product of two fields is defined, as usual, by
\[
: \phi (y) \rho (z) : = \phi (y)_+
\rho (z) - \rho (z) \phi (y)_-.
\]
Then one has (using Wick's formula) ($\lambda=\pm$):
\begin{equation}
  \label{leur:2.6}
\begin{aligned}
{}& [\alpha_k,\psi^{\lambda} (z),] =\lambda z^k\psi^\lambda(z),\\
{}& [\alpha_k,Y^{\lambda} (z),] = 2\lambda z^kY^\lambda(z),\\
{}& [L^\nu_k,\psi^\lambda(z)] =
\left(z^{k+1}\partial_z+\left(\lambda\left(\nu-\frac{1}{2}\right)+\frac{1}{2}\right)(k+1)z^k\right)
\psi^\lambda(z),\\
{}& [L^\nu_k,Y^\lambda(z)] =
\left(z^{k+1}\partial_z+\left(2\lambda\left(\nu-\frac{1}{2}\right)+2\right)(k+1)z^k\right)
Y^\lambda(z),\\
{}& [L^\nu_k, \alpha(z)] =\left(z^{k+1}\partial_z+(k+1)z^k\right)\alpha(z)+\left(\nu
-\frac{1}{2}\right)(k+1)kz^{k-1},\\
{}& [\alpha_k,\alpha(z)]=kz^{k-1},\\
{}& [L^\nu_k,L^\nu_\ell]=(k-\ell)L^\nu_{k+\ell}
+\delta_{k,-\ell}\frac{k^3-k}{12}c_\nu,
\end{aligned}
\end{equation}
where $c_\nu=-12\nu^2+12\nu-2$
and $\psi_0$ commutes with all these bosonic operators.

Thus, the $\alpha_k$, $L^\nu_k$ form the oscillator algebra, respectively
Virasoro algebra and
\begin{equation}
\label{leur:vac}
\begin{aligned}
{}& \alpha_k |0\rangle  = 0\quad &&\mbox{for}\ \ k\geq 0,\\
{}&L^\nu_k|0\rangle = 0\quad && \mbox{for}\ \ k\geq -1,\\
{}&Y_k^\pm|0\rangle = 0\quad && \mbox{for}\ \ k\geq 0.
\end{aligned}
\end{equation}

The operator  $\alpha_0$ is diagonalizable in $F$, with eigenspaces the
{\it charge sectors} $F_k$, i.e. $\alpha_0 f_k = kf_k$ for $f_k \in
F_k$. For that reason we call $\alpha_0$ the {\it charge operator}.

In order to express the fermionic fields $\psi^{\lambda} (z)$ in terms
of the oscillator algebra, we need an additional operator $Q$ on $F$
defined by
\[
Q |0\rangle  = \psi^+_{-\frac{1}{2}} |0\rangle , \qquad Q\psi^+_k =
\psi^{\pm}_{k\mp 1} Q,\qquad Q\psi_0=-\psi_0Q.
\]

\begin{proposition}  \label{leur:t2.1}
{\bf (\cite{leur:DJKM1,leur:JM,leur:K2})}
\[
\begin{aligned}
{}& \psi^{\pm} (z) = Q^{\pm 1} z^{\pm \alpha_0} \exp \left(\mp \sum_{k<0}
\frac{\alpha_k}{k} z^{-k}\right) \exp \left(\mp \sum_{k>0} \frac{\alpha_k}{k}
z^{-k}\right),\\[1mm]
{}& Y^{\pm} (z) = Q^{\pm 2} z^{\pm 2\alpha_0} \exp \left(\mp2 \sum_{k<0}
\frac{\alpha_k}{k} z^{-k}\right) \exp \left(\mp 2\sum_{k>0} \frac{\alpha_k}{k} z^{-k}\right).
\end{aligned}
\]
\end{proposition}

We now identify $F$ with the space $B=\mathbb{C} [q,q^{-1}, t_1, t_2 ,
\ldots ]$ via the vector space homorphism
\[
\sigma : F \tilde{\rightarrow} B
\]
given by
\[
\sigma \left(\alpha_{-m_1} \ldots \alpha_{-m_s} Q^k |0\rangle  \right) = m_1 m_2 \ldots
m_s
t_{m_1} t_{m_2} \ldots t_{m_s} q^k .
\]

The transported charge is as follows
\[
\mbox{charge} \left(p(t)q^k\right)=k,
\]
and the transported charge decomposition is
\[
\displaystyle
B=\bigoplus_{m\in \mathbb{Z}} B_m, \qquad \mbox{where}\quad B_m = \mathbb{C}
[t_1 , t_2 \ldots ]q^m .
\]

The transported operators $\psi_0$, $\alpha_m$ and $Q$ on $B$ are as
follows:
\begin{equation}
\label{leur:W}
\begin{aligned}
{}&\sigma\psi_0\sigma^{-1} =-\frac{1}{\sqrt 2}(-)^{q
\frac{\partial}{\partial q}}, \qquad & &
 W_{-m}=\sigma \alpha_{-m}\sigma^{-1} = mt_m,\\
{}& W_m=\sigma \alpha_m \sigma^{-1} = \frac{\partial}{\partial t_m}, \qquad
&& W_0=\sigma \alpha_0 \sigma^{-1} = q \frac{\partial}{\partial q},\\
{} &\sigma Q \sigma^{-1}= q.
\end{aligned}
\end{equation}

Hence
\[
\begin{aligned}
{}& \sigma\psi^{\pm} (z) \sigma^{-1} = q^{\pm 1} z^{\pm q\frac{\partial}{\partial
    q}} e^{\pm \xi (t,z)} e^{\mp \eta(t,z)},\\[1mm]
{}& \sigma Y^{\pm} (z) \sigma^{-1} = q^{\pm 2} z^{\pm 2q\frac{\partial}{\partial
    q}} e^{\pm 2\xi (t,z)} e^{\mp 2\eta(t,z)},
\end{aligned}
\]
where $\xi (t,z)$ and $\eta(t,z)$ are defined in (\ref{leur:0.4}).
It is now straightforward to prove the following

\begin{lemma}\label{leur:L1}
For $y>z$ one has
\[
\begin{aligned}
{}& \sigma\psi^\lambda(y)\psi^\mu(z)\sigma^{-1}=
(y-z)^{\lambda\mu 1}q^{\lambda+\mu} y^{\lambda q\frac{\partial}{\partial q}}
z^{\mu q\frac{\partial}{\partial q}} e^{\lambda\xi(t,y)+\mu\xi(t,z)}
e^{-\lambda\eta(t,y)-\mu\eta(t,z)},\\[1mm]
{}& \sigma Y^\lambda(y)Y^\mu(z)\sigma^{-1}=
(y-z)^{\lambda\mu 4}q^{\lambda2+\mu2} y^{\lambda2 q\frac{\partial}{\partial q}}
z^{\mu 2q\frac{\partial}{\partial q}} e^{\lambda 2\xi(t,y)+\mu2\xi(t,z)}
e^{-\lambda 2\eta(t,y)-\mu 2\eta (t,z)}.
\end{aligned}
\]
\end{lemma}

Using the commutation relations (\ref{leur:2.2}) one immediately obtains the
following consequence of this lemma:

\begin{corollary}
\label{leur:C1}
For $z_0,z_1,\ldots,z_m\in\mathbb{R}$ distinct we have
\[
\begin{aligned}
{}& \sigma\psi^+(z_m)\psi^+(z_{m-1})\cdots
\psi^+(z_1)\psi^+(z_0)\sigma^{-1}\\[1mm]
&{} \qquad = q^{m+1}\Delta_{m+1}(z)\prod_{i=0}^m z_i^{ q\frac{\partial}{\partial q}}
e^{\sum\limits_{j=0}^m\xi(t,z_j)}e^{-\sum\limits_{j=0}^m\eta(t,z_j)},\\[1mm]
{}& \sigma Y^+(z_m)Y^+(z_{m-1})\cdots Y^+(z_1)Y^+(z_0)\sigma^{-1}\\[1mm]
{}& \qquad =q^{2m+2}\Delta_{m+1}^4(z)\prod_{i=0}^m z_i^{ 2q\frac{\partial}{\partial q}}
e^{\sum\limits_{j=0}^m2\xi(t,z_j)}e^{-\sum\limits_{j=0}^m2\eta(t,z_j)}.
\end{aligned}
\]
\end{corollary}

We will now use the boson-fermion correspondence to rewrite the BKP hierarchy.
Notice first that (\ref{leur:1.10}) is equivalent to
\begin{equation}
\label{leur:0.10}
\mbox{Res}\left(
\psi^+(z)\tau\otimes\psi^-(z)\tau
+\psi^-(z)\tau\otimes\psi^+(z)\tau\right )
=\frac{1}{2}\left (\tau\otimes\tau -\psi_0\tau\otimes\psi_0\tau \right ),
\end{equation}
where $\text{Res}\,\sum f_iz^i=f_{-1}$.
Now apply $\sigma$ to (\ref{leur:0.10}). Writing
$\sigma(\tau)=\sum\limits_{n\in\mathbb{Z}}\tau_n(t)q^n$, we obtain for all
$n,m\in\mathbb{Z}$ equation (\ref{leur:0.12})
as the coefficient of $q^n\otimes q^m$.

The Spin group elements which we consider in the rest of the paper
do not fit in the algebraic framework of this section.
However, we will  need them to describe the symmetric  and symplectic matrix
integrals as (generalized) tau-functions. Since all manipulations with vertex
operators, used there, are well defined and correct, it is clear what we mean
and we are doing.

\setcounter{equation}{0}
\section{Pfaffian tau-functions}

Let $F(z)$ (resp. $F(y,z)$) be a (skew-symmetric) weight function on
$\mathbb{R}$ ($\mathbb{R}^2$) and let
\begin{equation}
\label{leur:P3.1}
\begin{aligned}
{}& (f,g)=\int_{\mathbb R}f(z)g(z)F(z)dz,\quad  \text{respectively}\\
{}& \langle f,g\rangle=\iint_{\mathbb{R}^2} f(y)g(z)F(y,z)dydz
\end{aligned}
\end{equation}
be the corresponding symmetric (skew-symmetric) inner product. Consider the
following deformation of these inner products, which we assume to be  the
time-dependent $t=(t_1,t_2,\cdots)$:
\begin{equation}
\label{leur:P3.2}
\begin{aligned}
{}& (f,g)_t=\int_{\mathbb R}f(z)g(z)e^{\xi(t,z)}F(z)dz,\\[1mm]
{}& \langle f,g\rangle_t=\iint_{\mathbb{R}^2}
f(y)g(z)e^{\xi(t,y)+\xi(t,z)}F(y,z)dydz,
\end{aligned}
\end{equation}
with $\xi(t,z)=\sum\limits_i t_iz^i$ as before. Notice that the inner products
appearing in Section~1 and~2 are special cases of (\ref{leur:P3.2}). Let
\[
\begin{aligned}
{}& \mu_i(t)=(1,z^i)_t\quad\text{and} \quad \mu_i=\mu_i(0)=(1,z^i),\\[1mm]
{} & \mu_{ij}(t)=\langle y^i,z^j\rangle_t=-\mu_{ji}(t)\quad \text{and}
\quad \mu_{ij}=\mu_{ij}(0)=\langle y^i,z^j\rangle=-\mu_{ji}
\end{aligned}
\]
be the moments. Consider the following generalized Spin group element (see
Section~3):
\[
G^\mu=\exp (g^\mu)\cdot f^\mu=
\exp\left (
\sum_{-\infty<j<i<0}\!\! \mu_{-i-\frac{1}{2},-j-\frac{1}{2}}\psi_j^+\psi_i^+\right
)\!\!\left (1+\sqrt 2 \sum_{k<0}\mu_{-k-\frac{1}{2}}\psi_0\psi^+_k\right ).
\]
We calculate its action on the vacuum vector $|0\rangle$, let
\begin{equation}
\label{leur:P3.a}
\begin{aligned}
\tau^F={}&\exp\left (
\sum_{-\infty<j<i<0} \mu_{-i-\frac{1}{2},-j-\frac{1}{2}}\psi_j^+\psi_i^+\right
)\left (1+\sqrt 2 \sum_{k<0}\mu_{-k-\frac{1}{2}}\psi_0\psi^+_k\right
)|0\rangle\\[1mm]
={}&\exp\left (\frac{1}{2}
\sum_{i,j<0} \mu_{-i-\frac{1}{2},-j-\frac{1}{2}}\psi_j^+\psi_i^+\right
)\left (1+ \sum_{k<0}\mu_{-k-\frac{1}{2}}\psi^+_k\right )|0\rangle\\[1mm]
={}&\sum_{n=0}^\infty \frac{1}{2^nn!}\left(
\sum_{i,j<0} \mu_{-i-\frac{1}{2},-j-\frac{1}{2}}\psi_j^+\psi_i^+\right
)^n\left (1+ \sum_{k<0}\mu_{-k-\frac{1}{2}}\psi^+_k\right )|0\rangle\\[1mm]
={}&\sum_{n=0}^\infty \frac{1}{2^nn!}\left(
\sum_{i,j<0}
\iint_{\mathbb{R}^2}y^{-i-\frac{1}{2}}z^{-j-\frac{1}{2}}
F(y,z)dydz\psi_j^+\psi_i^+\right )^n\\[1mm]
{}&\qquad\times\left (1+\int_{\mathbb R}\sum_{k<0}
w^{-k-\frac{1}{2}}\psi^+_kF(w)dw\right )|0\rangle
\end{aligned}
\end{equation}
\[
\begin{aligned}
\phantom{\tau^F}{}
={}&\sum_{n=0}^\infty \frac{1}{2^nn!}\left(
\iint_{\mathbb{R}^2}\psi^+(z)\psi^+(y)F(y,z)dydz\right )^n\left
(1+\int_{\mathbb R}\psi^+(w)F(w)dw\right )|0\rangle\\[1mm]
={}&\sum_{n=0}^\infty \frac{1}{2^nn!}\left(
 \int_{\mathbb{R}^{2n}}\psi^+(z_{2n-1})\psi^+(z_{2n-2})\cdots\psi^+(z_0)
\prod_{j=0}^{n-1}F(z_{2j},z_{2j+1})dz_{2j}dz_{2j+1}\right )\\[1mm]
{} &\qquad\times\left (1+\int_{\mathbb R}\psi^+(z_{2n})F(z_{2n})dz_{2n}\right )|0\rangle\\
={}&\sum_{m=0}^\infty \tau^F_{m}.
\end{aligned}
\]
Using the boson-fermion correspondence and corollary \ref{leur:C1} we rewrite this
as follows, let $\tau^F(t,q)=\sigma\left(\tau^F\right)$, then
\begin{equation}
\label{leur:P3.b}
\begin{aligned}
\tau^F(t,q)={}& \sum_{n=0}^\infty \frac{1}{2^nn!}\left(
\int_{\mathbb{R}^{2n}}\Delta_{2n}(z)
\prod_{j=0}^{n-1}e^{\xi(t,z_{2j})+\xi(t,z_{2j+1})}
F(z_{2j},z_{2j+1})dz_{2j}dz_{2j+1}\right)q^{2n}\\[1mm]
{}&\qquad +\sum_{n=0}^\infty \frac{1}{2^nn!}\Biggl(
\int_{\mathbb{R}^{2n+1}}\Delta_{2n+1}(z)e^{\xi(t,z_{2n})}F(z_{2n})dz_{2n}\\[1mm]
{}& \qquad  \times \prod_{j=0}^{n-1}e^{\xi(t,z_{2j})+\xi(t,z_{2j+1})}
F(z_{2j},z_{2j+1})dz_{2j}dz_{2j+1}\Biggr)q^{2n+1}\\[1mm]
={} &\sum_{m=0}^\infty \tau_{m}^F(t)q^{m}.
\end{aligned}\hspace{-10mm}
\end{equation}
Then $\tau_0^F(t)=1$,
\begin{equation}
\label{leur:P3.bb}
\begin{aligned}
\tau_{2n}^F(t)={}&\frac{1}{2^nn!}
\int_{\mathbb{R}^{2n}}\sum_{\sigma\in S_{2n}}\text{sg}(\sigma)
z_0^{\sigma(0)}z_1^{\sigma(1)}\cdots z_{2n-1}^{\sigma(2n-1)}\\[1mm]
{}&\qquad\times\prod_{j=0}^{n-1}e^{\xi(t,z_{2j})+\xi(t,z_{2j+1})}
F(z_{2j},z_{2j+1})dz_{2j}dz_{2j+1}\\[1mm]
={}&\frac{1}{2^nn!}
\sum_{\sigma\in
S_{2n}}\text{sg}(\sigma)\prod_{j=0}^{n-1}\mu_{\sigma(2j),\sigma(2j+1)}(t)\\[1mm]
={}& \text{Pf}((\mu_{k,\ell}(t))_{0\le k,\ell\le 2n-1})
\end{aligned}
\end{equation}
and
\begin{equation}
\label{leur:P3bc}
\begin{aligned}
\tau_{2n+1}^EF(t)={}&\frac{1}{2^nn!}
\int_{\mathbb{R}^{2n+1}}\sum_{\sigma\in S_{2n+1}}\text{sg}(\sigma)
z_0^{\sigma(0)}z_1^{\sigma(1)}\cdots z_{2n}^{\sigma(2n)}F(z_{2n})\\[1mm]
{}&\qquad\times\prod_{j=0}^{n-1}F(z_{2j},z_{2j+1})
\prod_{i=0}^{2n}e^{\xi(t,z_{i})}dz_{i}.
\end{aligned}
\end{equation}
A similar calculation as in (\ref{leur:P1.1b}) and (\ref{leur:P1.1c}) now shows that
$\tau_{2n+1}^F(t)=\text{Pf}(M_{2n+1}(t))$, where $M_{2n+1}(t)$ is given by
(\ref{leur:M}). Hence we have shown:

\begin{theorem} Let $\mu_i(t)$, $\mu_{ij}(t)$ be the moments of the symmetric,
respectively skew-symmetric
innerproduct $(\cdot,\cdot)_t)$ and $\langle\cdot,\cdot\rangle_t$, defined by
(\ref{leur:P3.2}). Let $M_{2n}(t)=(\mu_{ij}(t))_{0\le i,j\le 2n-1}$ and
\[
M_{2n+1}(t)=\left(
\begin{array}{ccc|c}
 & & &\mu_0(t)\\
 &(\mu_{ij}(t))_{0\le i,j\le 2n}& & \vdots\\
 & & & \vdots\\
 & & &\mu_{2n}(t)\\[1mm]
\hline
&&&\\[-3.5mm]
-\mu_0(t)&\cdots\ \cdots&-\mu_{2n}(t)&0
\end{array}\right),
\]
then
\[
\begin{aligned}
{}& \sigma\left (\exp\left (
\sum_{-\infty<j<i<0}
\mu_{-i-\frac{1}{2},-j-\frac{1}{2}}(0)\psi_j^+\psi_i^+\right )
\left (1+\sqrt 2 \sum_{k<0}\mu_{-k-\frac{1}{2}}(0)\psi_0\psi^+_k\right )
 |0\rangle \right )\\[1mm]
{}& \qquad =\exp \left
(\frac{1}{2}\iint_{\mathbb{R}^2}X(t,y,z)F(y,z)dydz\right )
\left( 1+\int_{\mathbb{R}}X(t,w)F(w)dw\right )\cdot 1\\[1mm]
{} & \qquad =\sum_{m=0}^\infty \text{\rm Pf}(M_m)(t)q^{m},
\end{aligned}
\]
where $X(t,y,z)$ and $X(t,w)$ are given by (\ref{leur:X}).
\end{theorem}

Since $\tau^F$ is an element in the spin group orbit of the vacuum vector, one
has the following consequence:
\begin{corollary}
The
Pfaffian tau-functions $\tau_{m}^F(t)=\text{\rm Pf}(M_m(t))$,
$m\in\mathbb Z$, satisfy the BKP hierarchy (\ref{leur:0.12}).
\end{corollary}

Since $f^\mu$ and $\exp (g^\mu)$ commute, we find that
\[
\tau_{2n+1}^F=\sqrt 2 \sum_{k<0}\mu_{-k-\frac{1}{2}}\psi_0\psi^+_k\tau^F_{2n}.
\]
Hence,
\[
\tau_{2n+1}^F(t)=\int_{\mathbb{R}}w^{2n}e^{\xi(t,w)}e^{-\eta(t,w)}
\tau_{2n}^F(t)F(w)dw
\]
and $\tau_{2n+1}^F(t)$ is completely determined by $\tau_{2n}^F(t)$

\setcounter{equation}{0}
\section{Virasoro constraints}

In this section we give a representation theoretical proof of the Virasoro
constraints for the symmetric matrix integrals obtained by Adler and van
Moerbeke in~\cite{leur:Pfaf}. Our proof is along the lines of \cite{leur:AvMspec}.

As before, consider the matrix integral over symmetric
matrices
\[
\hat\tau_{m}^E(t)=m!c_{m}\tau_{m}^E(t)=\int_{{\cal S}_{m}(E)}
e^{{\rm Tr}\left(V(Z)+\sum\limits_1^{\infty}t_iZ^i\right)}dZ,
\]
integrated over the space ${\cal S}_{m}(E)$ of symmetric matrices
with spectrum in $E\subset\mathbb R$, where
\begin{equation}
\label{leur:P4.0}
E=\text{disjoint union}
\bigcup^r_{i=1}[c_{2i-1},c_{2i}]
\end{equation}
and we assume that the potential $V$ satisfies
\begin{equation}
\label{leur:P4.2}
V'(z)=\frac{g(z)}{f(z)}=\frac{\sum\limits_0^{\infty}b_i z^i}{\sum\limits_0^{\infty}a_i z^i},
\end{equation}
with $e^{V(z)}$ decaying to $0$ fast enough at the boundary of its
support.
Let
\begin{equation}
\label{leur:P4.2a}
\begin{aligned}
{} & F(z)=e^{V(z)}I_E(z),\\[1mm]
{}& F(y,z)=e^{V(y)+V(z)} \text{sg}(z-y))I_E(y)I_E(z),
\end{aligned}
\end{equation}
be the corresponding weight functions.
Comparing (\ref{leur:P1.1}), (\ref{leur:P3.a}) and (\ref{leur:P3.b}) we find that
\begin{equation}
\label{leur:P4.1}
\begin{aligned}
{}& \hat\tau_{2n}^E(t)=\frac{(2n)!c_{2n}}{2^nn!} \sigma
\left(\left(
\iint_{\mathbb{R}^2}\psi^+(z)\psi^+(y)F(y,z)dydz\right )^n|0\rangle\right),\\[1mm]
{}& \hat\tau_{2n+1}^E(t)=\frac{(2n)!c_{2n}}{2^nn!} \sigma
\left(\left(\iint_{\mathbb{R}^2}\psi^+(z)\psi^+(y)F(y,z)dydz\right )^n\right.\\[1mm]
{}& \phantom{\hat\tau_{2n+1}^E(t)={}} \left. \times \left( \int_{\mathbb R}\psi^+(w)F(w)dw\right )|0\rangle\right ).
\end{aligned}
\end{equation}

Consider the Virasoro algebra (see (\ref{leur:2.3})) $L_k:=L_k^1$,
$k\in\mathbb{Z}$, with central charge $-2$. {}From the commutation relations
(\ref{leur:2.6})
we deduce
\begin{equation}
\label{leur:P4.3}
[L_k,\psi^+(z)]=\partial_z(z^{k+1}\psi^+(z)).
\end{equation}
Hence,
\begin{equation}\arraycolsep=0em
\label{leur:P4.4}
\begin{aligned}
{}& \left[ \,\sum_{\ell=0}^\infty  a_{\ell} L_{k+\ell},
\iint_{\mathbb{R}^2}\psi^+(z)\psi^+(y)F(y,z)dydz \right]\\[1mm]
{}& \qquad {} =\sum_{\ell=0}^\infty a_{\ell} \iint_{\mathbb{R}^2}\left[L_{k+\ell},
\psi^+(z)\psi^+(y)F(y,z)dydz\right]\\[1mm]
{} & \qquad {} = \sum_{\ell=0}^\infty a_{\ell}
\iint_{\mathbb{R}^2}F(y,z)\left(\partial_yy^{k+\ell+1}+\partial_zz^{k+\ell+1}\right)
(\psi^+(z)\psi^+(y))dydz\\[1mm]
{}& \qquad {}=\sum_{\ell=0}^\infty a_{\ell}
\iint_{E^2}\left(\partial_yy^{k+\ell+1}+\partial_zz^{k+\ell+1}\right)\psi^+(z)
\psi^+(y)e^{V(y)+V(z)}\text{sg}(z-y)dydz
\end{aligned}\hspace{-5mm}
\end{equation}
\[\arraycolsep=0em
\begin{aligned}
{}& {}+\sum_{\ell=0}^\infty a_{\ell}
\iint_{{E}^2}\left(y^{k+\ell+1}V'(y)+z^{k+\ell+1}V'(z)\right)\psi^+(z)
\psi^+(y)e^{V(y)+V(z)}\text{sg}(y-z)dydz\\[1mm]
{}& {}+2\sum_{\ell=0}^\infty a_{\ell}
\iint_{{E}^2}\left(z^{k+\ell+1}\delta(z-y)-y^{k+\ell+1}\delta(z-y)\right)
\psi^+(z)\psi^+(y)e^{V(y)+V(z)}dydz,
\end{aligned}
\]
where we have used $\partial_z\text{sg}(z-y)=2\delta(z-y)$.
The last term on the right-hand side of (\ref{leur:P4.4}) is equal to 0, the first
to
\[
\sum_{\ell=0}^\infty a_{\ell} \sum_{i=1}^{2r} c_i^{k+\ell+1}\partial_{c_i}\left
(\iint_{\mathbb{R}^2}\psi^+(z)\psi^+(y)F(y,z)dydz\right )
\]
and the second to
\[
\begin{aligned}
{}& {}-\sum_{\ell=0}^\infty
b_\ell\iint_{E^2}\left(y^{k+\ell+1}+z^{k+\ell+1}\right)
\psi^+(z)\psi^+(y)e^{V(y)+V(z)}\text{sg}(z-y)dydz\\[1mm]
{}& \qquad {} =-\sum_{\ell=0}^\infty
b_\ell\iint_{\mathbb{R}^2}\left[\alpha_{k+\ell+1},\psi^+(z)\psi^+(y)\right]F(y,z)dydz\\[1mm]
{}& \qquad {}= -\sum_{\ell=0}^\infty
b_\ell\left[\alpha_{k+\ell+1},\iint_{\mathbb{R}^2}\psi^+(z)\psi^+(y)F(y,z)dydz\right].
\end{aligned}
\]
{}From which we conclude that
\[
\left[\sum_{\ell=0}^\infty \left (a_{\ell}\sum_{i=1}^{2r}
c_i^{k+\ell+1}\partial_{c_i}
-a_{\ell} L_{k+\ell}-b_\ell \alpha_{k+\ell+1}\right
),\iint_{\mathbb{R}^2}\psi^+(z)\psi^+(y)F(y,z)dydz\right]=0.
\]
An analogous calculation also shows that
\[
\left[\sum_{\ell=0}^\infty \left (a_{\ell}\sum_{i=1}^{2r}
c_i^{k+\ell+1}\partial_{c_i}
-a_{\ell} L_{k+\ell}-b_\ell \alpha_{k+\ell+1}\right
),\int_{\mathbb{R}}\psi^+(w)F(w)dw\right]=0.
\]

Using the action of the Virasoro and oscillator algebra on the vacuum vector
(\ref{leur:vac}), one has that for all $k\ge -1$ and $m\ge 0$:
\begin{equation}
\label{leur:P4.5}
\sum_{\ell=0}^\infty \left (a_{\ell}\sum_{i=1}^{2r}
c_i^{k+\ell+1}\partial_{c_i}
-a_{\ell} L_{k+\ell}-b_\ell \alpha_{k+\ell+1}\right )\left(\tau_{m}^F\right)=0.
\end{equation}
Now recall from (\ref{leur:W}) that $W_k=-kt_{-k},\ =q\frac{\partial}{\partial q},\
=\frac{\partial}{\partial t_k}$ for $k<0,\ k=0,\ k>0$, respectively. Define
\[
W_k^{(2)}=
\frac{1}{2}
\delta_{\left[\frac{k}{2}\right],\frac{k}{2}}W_{\frac{k}{2}}^2
+\sum_{i>\left[\frac{k}{2}\right]}W_{k-i}W_i+\frac{k+1}{2} W_k,
\]
here $\left[\frac{N}{2}\right]=\text{Entier}\left(\frac{N}{2}\right)$, and let as in Section 1
\[
\tau^E(t,q)=\sum_{m=0}^\infty \frac{\hat\tau_{m}^E(t)}{m!c_{m}}q^{m},
\]
then using the boson-fermion correspondence we have the following result of
Adler and van Moerbeke~\cite{leur:Pfaf}:
\begin{proposition}
For all $k\ge -1$ and all $m\ge 0$,
\[
\begin{aligned}
{}& \sum_{\ell=0}^\infty \left (a_{\ell}\sum_{i=1}^{2r}
c_i^{k+\ell+1}\partial_{c_i}
-a_{\ell} W^{(2)}_{k+\ell}-b_\ell W_{k+\ell+1}\right )\left
(\hat\tau_{m}^E(t)q^{m}\right )=0,\\[2mm]
{}& \sum_{\ell=0}^\infty \left (a_{\ell}\sum_{i=1}^{2r}
c_i^{k+\ell+1}\partial_{c_i}
-a_{\ell} W^{(2)}_{k+\ell}-b_\ell W_{k+\ell+1}\right )\left (\tau^E(t,q)\right)=0.
\end{aligned}
\]
\end{proposition}

\setcounter{equation}{0}

\section{Fay identity of the BKP hierarchy}

In this section we will prove the generalized Fay identity. We will use this
identity to prove a recursion relation between certain pfaffian wave functions,
which we will define in the next section.
\begin{proposition}
\label{leur:P5} {\bf (Fay identity)}
Let $t-t'=\sum\limits_{k=1}^p[z_k]-\sum\limits_{\ell=1}^q[y_\ell]$, where
$[z]=\left(\frac{z}{1},\frac{z}{2},\frac{z}{3},\ldots\right)$, then
\[
\begin{aligned}
{}&\frac{1}{2}\left (1-(-1)^{n+m}\right )\tau_n(t)\tau_m(t')\\
{} &\qquad {}= \sum_{k=1}^pz_k^{m-n+p-q}\tau_{n-1}(t-[z_k])\tau_{m+1}(t'+[z_k])
\frac{\prod\limits_{i=1}^qz_k-y_i}{\prod\limits_{j=1,j\ne k}^pz_k-z_j}\\
{} & \qquad {}+\sum_{r=0}^\infty
\frac{\delta_{n-m-p+q-1,r}}{r!}\left(\frac{\partial}{\partial z}\right)^r
\left(\tau_{n-1}(t-[z])\tau_{m+1}(t'+[z])
\frac{\prod\limits_{i=1}^qz-y_i}{\prod\limits_{j=1}^pz-z_j}\right )\Biggr |_{z=0}\\
{} &\qquad{} +\sum_{\ell=1}^qy_\ell^{n-m+q-p}\tau_{n+1}(t+[y_\ell])\tau_{m-1}(t'-[y_\ell])
\frac{\prod\limits_{j=1}^py_\ell -z_p}{\prod\limits_{i=1,i\ne \ell}^qy_\ell-y_i}\\
{} &\qquad {}+\sum_{r=0}^\infty
\frac{\delta_{m-n+p-q-1,r}}{r!}\left(\frac{\partial}{\partial z}\right)^r
\left(\tau_{n+1}(t+[z])\tau_{m-1}(t'-[z])
\frac{\prod\limits_{j=1}^pz-z_j}{\prod\limits_{i=1}^qz-y_i}\right )\Biggr|_{z=0}.
\end{aligned}
\]
\end{proposition}

\noindent
{\bf Proof.} Notice that equation (\ref{leur:0.12}) is equal to
\begin{equation}
\label{leur:5.12}
\begin{aligned}
{}& \frac{1}{2}\left (1-(-1)^{n+m}\right )\tau_n(t)\tau_m(t')\\
{}& \qquad {}= \frac{1}{2\pi i}\oint_{z=0}\Big(
z^{n-m}\tau_{n-1}(t-[z])\tau_{m+1}(t'+[z])e^{\xi(t,z^{-1})-\xi(t',z^{-1})}\\
{}& \qquad {}+ z^{m-n}\tau_{n+1}(t+[z])\tau_{m-1}(t'[z])e^{\xi(t',z^{-1})-\xi(t,z^{-1})}
\Big)dz,
\end{aligned}
\end{equation}
where the integration is taken along a contour around $z=0$ and all $z=y_k$ and
$z=z_\ell$.
Since $t-t'=\sum\limits_{k=1}^p[z_k]-\sum\limits_{\ell=1}^q[y_\ell]$,
we find that
\[
e^{\xi(t,z^{-1})-\xi(t',z^{-1})}=
z^{p-q}\frac{\prod\limits_{i=1}^qz-y_i}{\prod\limits_{j=1}^pz-z_j}.
\]
Hence the right-hand side of (\ref{leur:5.12}) is equal to
\[
\begin{aligned}
{}& \frac{1}{2\pi i}\oint_{z=0}\left(
z^{m-n+p-q}\tau_{n-1}(t-[z])\tau_{m+1}(t'+[z])
\frac{\prod\limits_{i=1}^qz-y_i}{\prod\limits_{j=1}^pz-z_j}\right.\\[1mm]
{}& \qquad \left. {}+ z^{n-m+q-p}\tau_{n+1}(t+[z])\tau_{m-1}(t'-[z])
\frac{\prod\limits_{j=1}^pz-z_j}{\prod\limits_{i=1}^qz-y_i}\right)dz.
\end{aligned}
\]
Now using the fact that the integrand has poles only at $z=0$ and all $z=y_k$
and $z=z_\ell$, we obtain the desired result.\hfill \rule{3mm}{3mm}

\medskip

As a particular case of  Proposition \ref{leur:P5} we take $n=N+1$, $m=N-2$, $p=1$,
$z_1=u^{-1}$, $q=0$, $t'=t-[u^{-1}]$ and observe that
\[
\frac{\partial \tau(t\pm [z])}{\partial z}\Big |_{z=0}=
\pm\frac{\partial \tau(t\pm [z])}{\partial t_1}.
\]
We thus obtain
\begin{corollary}
\label{leur:C5}
\[
\begin{aligned}
{} & \tau_{N-1}(t)\tau_N\left(t-\left[u^{-1}\right]\right)=\tau_N(t)\tau_{N-1}\left(t-\left[u^{-1}\right]
\right)\\[2mm]
{}& \qquad {}+u^{-1}\left(
\tau_N(t)\frac{\partial \tau_{N-1}\left(t-\left[u^{-1}\right]\right)}{\partial t_1}
- \tau_{N-1}\left(t-\left[u^{-1}\right]\right)\frac{\partial\tau_N(t)}{\partial t_1}\right)\\[2mm]
{}& \qquad {}+u^{-2}\tau_{N+1}(t)\tau_{N-2}\left(t-\left[u^{-1}\right]\right).
\end{aligned}
\]
\end{corollary}

\setcounter{equation}{0}

\section{Wave functions and skew orthogonal polynomials}
Introduce the following BKP wave functions:
\begin{equation}
\Psi_n(t,z)=z^nP_n(t,z)e^{\xi(t,z)}:=
z^n\frac{\tau_n\left(t-\left[z^{-1}\right]\right)}{\tau_{n+1}(t)}.
\end{equation}
These are different wave functions than the one described in~\cite{leur:KvdL}
and~\cite{leur:Pfaf}. Let $h_n(t)=
\frac{(\tau_{n+1}(t))^2}{\tau_n(t)\tau_{n+2}(t)}$, using Corollary~\ref{leur:C5},
one easily deduces the following
recursion relation:
\begin{equation}
\label{leur:p5.1}
\Psi_n(t,z)= h_{n-1}(t)\frac{\partial\Psi_{n-1}(t,z)}{\partial t_1}
+\Psi_{n-2}(t,z).
\end{equation}
{}From now on let $\tau_n (t)=\tau_n^F(t)=\text{Pf}(M_n(t))$ as in Section~4.
It was shown in~\cite{leur:AHvM} that
\[
\begin{aligned}
{} & \tilde P_{2n}(t,z):=P_{2n}(t,z)\\[2mm]
{} & {}= \frac{1}{\tau_{2n+1}(t)}\text{Pf}\left(
\begin{array}{ccc|c}
 & & &1\\
 &(\mu_{ij}(t))_{0\le i,j\le 2n}& & z\\
 & & & \vdots\\
 & & &z^{2n}\\[1mm]
\hline
&&&\\[-3.5mm]
-1\ -z&\cdots\  \cdots&-z^{2n}&0
\end{array}\right),
\qquad\text{and}
\end{aligned}
\]
\[\hspace*{-10mm}
\begin{aligned}
{}& \tilde P_{2n+1}(t,z):=
\frac{h_{2n}(t)}{\tau_{2n+1}(t)}
(z+\frac{\partial}
{\partial t_1})\tau_{2n+1}P_{2n}(t,z)\\[2mm]
{} & =\frac{h_{2n}(t)}{\tau_{2n+1}(t)}
\text{Pf}\left(
\begin{array}{ccc|cc}
  & & &1&\mu_{0,2n+1}(t)\\
  &\!\!\!(\mu_{ij}(t))_{0\le i,j\le 2n-1}\!\!\!& & z&\mu_{1,2n+1}(t)\\
  & & & \vdots&\vdots\\
  & & &z^{2n-1}&\mu_{2n-1,2n+1}(t)\\[1mm]
\hline
&&&&\\[-3.5mm]
-1&\cdots \ \cdots&-z^{2n-1}&0&-z^{2n+1}\\
-\mu_{0,2n+1}(t)&\cdots\ \cdots&-\mu_{2n-1,2n+1}(t)&z^{2n+1}&0
\end{array}\right)\!,
\end{aligned}
\]
form a set skew orthonormal polynomials $\{ \tilde P_n(t,z)\}_{n\ge 0}$ with
respect to the time-dependent skew symmetric innerproduct $\langle
\cdot,\cdot\rangle_t$, defined in (\ref{leur:P3.2}), i.e.
\[
\begin{aligned}
{}& \langle \tilde P_{2m},\tilde P_{2n}\rangle_t=
\langle \tilde P_{2m+1},\tilde P_{2n+1}\rangle_t=0,\\[1mm]
{}& \langle \tilde P_{2m},\tilde P_{2n+1}\rangle_t=
-\langle \tilde P_{2n+1},\tilde P_{2m}\rangle_t=\delta_{nm}.
\end{aligned}
\]
Now using (\ref{leur:p5.1}), one sees that
\[
\tilde P_{2n+1}(t,z)-{h_{2n}(t)}
\frac{\partial\log\tau_{2n+1}(t)}
{\partial t_1}\tilde P_{2n}(t,z)= P_{2n+1}(t)- P_{2n-1}(t).
\]
Let $Q_{2n}(t,z)=P_{2n}(t,z)$ and $Q_{2n+1}(t,z)=P_{2n+1}(t)- P_{2n-1}(t)$,
we thus have the following result:

\begin{proposition}
\label{leur:P6}
Let  $\tau_n (t)=\tau_n^F(t)=\text{\rm Pf}(M_n(t))$, then the polynomials
 \[
\begin{aligned}
{}& Q_{2n}(t,z):=z^{2n}\frac{\tau_{2n}\left(t-\left[z^{-1}\right]\right)}{\tau_{2n+1}(t)}\\[2mm]
{}& {} =\frac{1}{\tau_{2n+1}(t)}\text{\rm Pf}\left(
\begin{array}{ccc|c}
 & & &1\\
 &(\mu_{ij}(t))_{0\le i,j\le 2n}& & z\\
 & & & \vdots\\
 & & &z^{2n}\\[1mm]
\hline
&&&\\[-3.5mm]
-1\ -z&\cdots\  \cdots&-z^{2n}&0
\end{array}\right),
\quad\text{and}
\end{aligned}
\]
\[\hspace*{-10mm}
\begin{aligned}
{}& Q_{2n+1}(t,z):=z^{2n+1}\frac{\tau_{2n+1}\left(t-\left[z^{-1}\right]\right)}{\tau_{2n+2}(t)}-
z^{2n-1}\frac{\tau_{2n-1}\left(t-\left[z^{-1}\right]\right)}{\tau_{2n}(t)}\\[2mm]
{}& =\frac{1}{\tau_{2n}(t)\tau_{2n+2}(t)}
\text{\rm Pf}\left(\!
\begin{array}{ccc|cc}
  & & &1&N_{0,2n}(t)\\
  &\!\!\!\!(\mu_{ij}(t))_{0\le i,j\le 2n-1}\!\!\!\!\!& & z&N_{1,2n}(t)\\
  & & & \vdots&\vdots\\
  & & &z^{2n-1}&\!\!N_{2n-1,2n}(t)\\[1mm]
\hline
&&&&\\[-3.5mm]
-1&\cdots \ \cdots& -z^{2n-1}&0&-N_{2n}(t,z)
\\
-N_{0,2n}(t)&\cdots\ \cdots&-N_{2n-1,2n}(t)\! &\! N_{2n}(t,z)&0
\end{array}\!\right)\!,
\end{aligned}
\]
where
\[
\begin{aligned}
{}& N_{j,2n}(t)=\mu_{j,2n}(t)\frac{\partial \tau_{2n+1}(t)}{\partial
t_1}+\mu_{j,2n+1}(t)\tau_{2n+1}(t),\qquad \text{for} \quad 1\le j< 2n,\\[3mm]
{}& N_{2n}(t,z)=z^{2n}\frac{\partial \tau_{2n+1}(t)}{\partial
t_1}+z^{2n+1}\tau_{2n+1}(t)
\end{aligned}
\]
form a set skew orthonormal polynomials with respect to the time-dependent
skew-sym\-met\-ric innerproduct $\langle \cdot,\cdot\rangle_t$. Moreover,
\[
\begin{aligned}
{} & Q_{2n+1}(t,z)={h_{2n}(t)} \left (z+\frac{\partial}
{\partial t_1}\right )Q_{2n}(t,z)\\[2mm]
{}& Q_{2n}(t,z)=\sum_{k=1}^n\left ( \sum_{\ell=1}^k
{h_{2n-2\ell+1}}\right )\left (z+\frac{\partial}
{\partial t_1}\right )Q_{2n-2k+1}(t,z).
\end{aligned}
\]
\end{proposition}

\setcounter{equation}{0}

\section{Symplectic matrix integrals}

In this section we will treat the symplectic matrix integrals, i.e.,
integrals of the form:
\[
 \int_{{\cal T}_{2n}(E)}
 e^{2\text{Tr}\left(V(Z)+\sum\limits_1^{\infty} t_i Z^i\right )} dZ,
 \]
where $dZ$ denotes the Haar measure
\[
dZ=\prod^N_1 dZ_k \prod_{k \leq \ell}
 dZ_{k \ell}^{(0)}\overline {dZ_{k \ell}^{(0)}}
   dZ_{k \ell}^{(1)} \overline {dZ_{k \ell}^{(1)}},
 \]
 on the space
${\cal
T}_{2N}(E)$ of self-dual $N\times N$ Hermitean matrices with quaternionic
entries and spectrum in $E\subset \mathbb R$; these particular  matrices can be
realized as the space of $2N
\times 2N$ matrices with entries $Z^{(i)}_{
\ell k}\in \mathbb C$
\[
{\cal T}_{2N}=\left\{ Z=(Z_{k \ell})_{1\leq k ,\ell \leq N} |
Z_{k \ell}=\left(\begin{array}{cc}Z^{(0)}_{k \ell}&Z^{(1)}_{k \ell}\vspace{1mm}\\
        -\overline Z^{(1)}_{k \ell}&\overline Z^{(0)}_{k \ell}
\end{array}\right ) \
        \mbox{with} \ \ Z_{ \ell k}=\overline Z_{k \ell}^{\top}
\right\},
\]
It is shown in \cite{leur:Pfaf2} that
\[
\begin{aligned}
\hat\tau_{2n}^E(t)={} &
 \int_{{\cal T}_{2n}(E)}
 e^{2\text{Tr}\left(V(Z)+\sum\limits_1^{\infty} t_i Z^i \right)} dZ\\[2mm]
={}&
\int_{E^n}\Delta^4_n(z)
  \prod_{i=1}^{n}\left(e^{\left(V(z)+\sum\limits_1^{\infty} t_i z^i\right) } dz_i
   \right)\\[2mm]
={} & n!\text{Pf}((\mu_{ij}(t))_{0\le i,j\le 2n-1}),
\end{aligned}
\]
where the $\mu_{ij}(t)$ are the moments of
the skew-symmetric innerproduct
\begin{equation}
\label{leur:7.1}
\langle f,g\rangle_t=\int_{\mathbb{R}}
\left(\frac{\partial f(z)}{\partial z}g(z)-\frac{\partial g(z)}{\partial
z}f(z)\right )e^{2\xi(t,z)}F(z)dz,
\end{equation}
and $F(z)=e^{2V(z)}I_E(z)$.
We will now show that the generating series
\[
\tau^E(t,q)=\sum_{n=0}^\infty \tau_{2n}^E(t)q^{2n},\qquad \text{with}\quad
\hat\tau_{2n}^E(t)=n!\tau_{2n}^E(t)
\]
of these Pfaffians is again an element in the Spin group orbit of
the vacuum vector.

Let $F(z)$  be a weight function on
$\mathbb{R}$ and introduce
the following symmetric, respectively skew-symmetric inner product:
\begin{equation}
\label{leur:7.2}
\begin{aligned}
{}& (f,g)=\int_{\mathbb R}f(z)g(z)F(z)dz,\quad \text{respectively}\\[1mm]
{}& \langle f,g\rangle=\int_{\mathbb R}\left(\frac{\partial f(z)}{\partial
z}g(z)-\frac{\partial g(z)}{\partial z}f(z)\right )F(z)dz
\end{aligned}
\end{equation}
with moments $\mu_{i}=(z^i,1)$ respectively $\mu_{ij}=\langle z^i,z^j\rangle$.
Notice that
$\langle \cdot ,\cdot \rangle_t$, given by (\ref{leur:7.1}), is a
time-dependent $t=(t_1,t_2,\cdots)$
deformation of $\langle \cdot,\cdot\rangle$. In a similar way is
\begin{equation}
\label{leur:7.3}
(f,g)_t=\int_{\mathbb R}f(z)g(z)e^{2\xi(t,z)}F(z)dz,
\end{equation}
a the time-dependent deformation of $(\cdot,\cdot)$.

To describe this symplectic case, we consider the generalized Spin group
element
\[
\begin{aligned}
H^\mu={}& \exp (h^\mu)=\exp\left(
\sum_{j<i<0}\mu_{-j-\frac{1}{2},-i-\frac{1}{2}}\psi_j^+\psi_i^+\right)\\[1mm]
={}& \exp\left(
\sum_{j<i<0}(i-j)\mu_{-i-j-2}\psi_j^+\psi_i^+\right)
=\exp\left(
\sum_{i,j<0}\left(-j-\frac{1}{2}\right )\mu_{-i-j-2}\psi_j^+\psi_i^+\right)
\end{aligned}
\]
and calculate its action on the vacuum vector:
\begin{equation}
\label{leur:7.4}
\begin{aligned}
\tau^F={}&\exp\left(
\sum_{j<i<0}\mu_{-j-\frac{1}{2},-i-\frac{1}{2}}\psi_j^+\psi_i^+\right)|0\rangle\\[1mm]
={} &\sum_{n=0}^\infty\frac{1}{n!}\left(
\sum_{i,j<0}\left (-j-\frac{1}{2}\right )\mu_{-i-j-2}\psi_j^+\psi_i^+
\right)^n|0\rangle\\[1mm]
={}&\sum_{n=0}^\infty\frac{1}{n!}\left(\sum_{i,j<0}\int_{\mathbb R}
z^{-i-j-2}F(z)dz \left (-j-\frac{1}{2}\right )\psi_j^+\psi_i^+
\right)^n|0\rangle\\[1mm]
={}&\sum_{n=0}^\infty\frac{1}{n!}\left(\int_{\mathbb R}
Y^+(z)F(z)dz \right)^n|0\rangle\\
={}&\sum_{n=0}^\infty\frac{1}{n!}\int_{\mathbb R^n}
Y^+(z_{n-1})Y^+(z_{n-2})\cdots Y^+(z_{0})\prod_{j=0}^{n-1}
F(z_j)dz_j |0\rangle\\[1mm]
={}&\sum_{n=0}^\infty \tau_{2n}^F.
\end{aligned}
\end{equation}
Use again the boson-fermion correspondence and corollary \ref{leur:C1}. Let
$\tau^F(t,q)=\sigma(\tau^F)$, then
\begin{equation}
\label{leur:7.5}
\tau^F(t,q)=\sum_{n=0}^\infty\frac{1}{n!}\int_{\mathbb R^n}
\Delta^4_n(z)\prod_{j=0}^{n-1}e^{2\xi(t,z_j)}
F(z_j)dz_j q^{2n}
=\sum_{n=0}^\infty \tau_{2n}^F(t)q^{2n}
\end{equation}
Let as in Section 5 $E$ satisfy (\ref{leur:P4.0}) and assume that
$F(z)=e^{2V(z)I_E(z)}$, where $V(z)$ satis\-fies~(\ref{leur:P4.2}).
Let $L(z)=L^0(z)$ (c.f. (\ref{leur:2.3}) )
be the Virasoro field. Although this field has central charge $-2$,
note that
it is another Virasoro field
than the one one considers in the symmetric case.
In a similar way as Section~5 one shows that
\[
\left[\sum_{\ell=0}^\infty \left (a_{\ell}\sum_{i=1}^{2r}
c_i^{k+\ell+1}\partial_{c_i}
-a_{\ell} L_{k+\ell}-b_\ell \alpha_{k+\ell+1}\right
),\int_{\mathbb{R}}Y^+(w)F(w)dw\right]=0.
\]
Which leeds to the following result.

\begin{theorem}
Let $\mu_{ij}(t)$ be the moments of the skew-symmetric
inner product (see (\ref{leur:7.1}))
$\langle\cdot,\cdot\rangle_t$  with weight function $F(z)=e^{2V(z)}I_E(z)$
and let
\[
Y(t,z)=q^2z^{2q\frac{\partial}{\partial q}}\exp\left( 2\sum_{i=1}^\infty t_iz^i\right)
\exp\left( -2\sum_{i=1}^\infty \frac{z^{-i}}{i}\frac{\partial}{\partial t_i}\right),
\]
then
\[
\begin{aligned}
{}& \sum_{n=0}^\infty \frac{1}{n!}\hat\tau_{2n}^E(t)q^{2n}
=\sigma\left(\exp\left(
\sum_{j<i<0}\mu_{-j-\frac{1}{2},-i-\frac{1}{2}}(0)
\psi_j^+\psi_i^+\right)|0\rangle\right)\\[1mm]
{}& \qquad {}=\exp\left ( \int_{\mathbb R} Y(t,z)F(z)dz\right )\cdot 1.
\end{aligned}
\]
All $\tau_{2n}^E(t)$ satisfy the DKP
hierarchy, i.e., the BKP hierarchy but only for the even tau-functions.
Moreover, let
\[
W^{(2)}_k=\frac{1}{2}
\delta_{[\frac{k}{2}],\frac{k}{2}}W_{\frac{k}{2}}^2
+\sum_{i>[\frac{k}{2}]}W_{k-i}W_i-\frac{k+1}{2} W_k,
\]
then for all $k\ge -1$ and all $n\ge 0$,
\[
\sum_{\ell=0}^\infty \left (a_{\ell}\sum_{i=1}^{2r}
c_i^{k+\ell+1}\partial_{c_i}
-a_{\ell} W^{(2)}_{k+\ell}-b_\ell W_{k+\ell+1}\right
)\left(\hat\tau_{2n}^E(t)q^{2n}\right )=0.
\]
\end{theorem}

\setcounter{equation}{0}

\section{Consequences of the Virasoro constraints}

In this section we will describe some
consequences of the Virasoro constraints. First consider the following.
Let $\tau=g|0\rangle$, $u\in \text{\rm Ann}\; \tau$ and
let $A$ be an operator, such that $[A,V]\subset V$ and $A\tau=0$,
then
\[
0=Au\tau=[A,u]\tau+uA\tau=[A,u]\tau
\]
and thus $[A,u]\in \text{\rm Ann}\; \tau$.
The operators ($k\ge -1$)
\[
A_k:=
\sum_{\ell=0}^\infty \left (a_{\ell}\sum_{i=1}^{2r}
c_i^{k+\ell+1}\partial_{c_i}
-a_{\ell} L_{k+\ell}-b_\ell \alpha_{k+\ell+1}\right
)
\]
satisfy this condition for $\tau=\tau^F$ in both symmetric and
symplectic cases and for
$\tau^g:=\sum\limits_{m=0}^\infty \tau_{2m}^Fq^{2m}=\exp(g^\mu)|0\rangle$
in the symmetric case.

First, we consider the symmetric case. We calculate $ \text{Ann}\;\tau^F=T_{G^\mu}(\text{Ann}\;
|0\rangle)=T_{G^\mu}(U_0)$.
Since for all $k>0$ and all $\ell$
\[
T_{f^\mu}(\psi_k^-)=\psi_k^-+\sqrt
2\mu_{k-\frac{1}{2}}\psi_0-\mu_{k-\frac{1}{2}}\sum_{i<0}
\mu_{-i-\frac{1}{2}}\psi^+_i,
\]
$T_{f^\mu}(\psi_k^+)=\psi_k^+$, $[g^\mu ,\psi^+_\ell]=[g^\mu,\psi_0]=0$ and
$[g^\mu,\psi^-_k] = \sum\limits_{i<0}
\mu_{k-\frac{1}{2},-i-\frac{1}{2}}\psi^+_i$,
we find that
\[
\text{\rm Ann}\;\tau^F=\sum_{k>0}
\mathbb{C}\psi_k^+ +\mathbb{C}\Psi_k^-,
\]
with
\[
\Psi_k^-=
\psi_k^-+\sqrt 2\mu_{k-\frac{1}{2}}\psi_0+\sum_{i>0}
\left (\mu_{k-\frac{1}{2},i-\frac{1}{2}}
-\mu_{k-\frac{1}{2}}\mu_{i-\frac{1}{2}}\right )\psi^+_{-i}.
\]
In simmilar way one obtains
\[
\text{\rm Ann}\;\tau^g=\sum_{k>0}
\mathbb{C}\psi_k^+ +\mathbb{C}\Phi_k^-,
\]
with
\[
\Phi_k^-=
\psi_k^-+\sum_{i>0}
\mu_{k-\frac{1}{2},i-\frac{1}{2}}
\psi^+_{-i}
\]
In the symmetric case we have:
\[
[\alpha_k,\psi_i^\pm]=\pm \psi^\pm_{i+k},\qquad
[L_k,\psi_i^\pm]=-i-\frac{k}{2}\pm\frac{k+1}{2}.
\]
Since,
\[
[A_k,\psi_j^-]
=\sum_{\ell=0}\left(j+k+\ell+\frac{1}{2}\right)
a_\ell \psi_{j+k+\ell}^-+b_\ell \psi_{j+k+\ell+1},
\]
one has that
\begin{equation}
\label{leur:psi}
\begin{aligned}
{}[A_k,\Psi_j^-]
={}&\sum_{\ell=0}\left(j+k+\ell+\frac{1}{2}\right)
a_\ell \Psi_{j+k+\ell}^-+b_\ell \Psi_{j+k+\ell+1}\\
\ &{} +{} \ \text{part which contains only $\psi_i^+$'s with $i>0$}.
\end{aligned}
\end{equation}
Now comparing the coefficient on both sides of $\psi_0$, we deduce
the following equation for all $k\ge -1$:
\[
\sum_{\ell=0}^\infty a_\ell\sum_{p=1}^{2r} c_p^{k+\ell+1}
\frac{\partial \mu_j}{\partial c_p}=
\sum_{\ell=0}^\infty (j+k+\ell+1)a_\ell \mu_{j+k+\ell}
+b_\ell\mu_{j+k+\ell+1}.
\]
The same equation (\ref{leur:psi}) holds with $\Psi_j^-$ replaced by
$\Phi_j^-$. Now consider the coefficient of $\psi_{-i}^+$
for $i>0$ on both sides. This leads to:
\[
\begin{aligned}
\sum_{\ell=0}^\infty a_\ell\sum_{p=1}^{2r} c_p^{k+\ell+1}
\frac{\partial \mu_{j,i}}{\partial c_p}={}&
\sum_{\ell=0}^\infty a_\ell\left((i+k+\ell+1)\mu_{j,i+k+\ell}
+(j+k+\ell+1) \mu_{j+k+\ell,i}\right)\\[1mm]
{} & {}+
b_\ell\left(\mu_{j,i+k+\ell+1}
+ \mu_{j+k+\ell+1,i}\right).
\end{aligned}
\]

In the symplectic case we find:
\[
\text{\rm Ann}\; \tau^F=\sum_{k>0}
\mathbb{C}\psi_k^+ +\mathbb{C}\Theta_k^-,
\]
with
\[
\Theta_k^-=
\psi_k^-+\sum_{i>0}
(i-k)\mu_{k+i-2}\psi^+_{-i}.
\]
In this case
\[
[\alpha_k,\psi_i^\pm]=\pm \psi^\pm_{i+k},\qquad
[L_k,\psi_i^\pm]=-i-\frac{k}{2}\mp\frac{k+1}{2},
\]
which leads to the following result for the symplectic case:
\[
\sum_{\ell=0}^\infty a_\ell\sum_{p=1}^{2r} c_p^{k+\ell+1}
\frac{\partial \mu_j}{\partial c_p}=
\sum_{\ell=0}^\infty (j+k+\ell+1)a_\ell \mu_{j+k+\ell}
+2b_\ell\mu_{j+k+\ell+1}
\]
and
\[
\sum_{\ell=0}^\infty a_\ell\sum_{p=1}^{2r} c_p^{k+\ell+1}
\frac{\partial \mu_{j,i}}{\partial c_p}=
\sum_{\ell=0}^\infty a_\ell\left(i\mu_{j,i+k+\ell}
+j \mu_{j+k+\ell,i}\right)
 +
b_\ell\left(\mu_{j,i+k+\ell+1}
+ \mu_{j+k+\ell+1,i}\right).
\]

\label{Leur-lastpage}

\newpage

\end{document}